\documentclass[journal,10pt,letterpaper,final,twocolumn]{IEEEtran}%
\usepackage{amsmath}
\usepackage{amsfonts}
\usepackage{cite}
\usepackage{amssymb}
\usepackage{mathrsfs}
\usepackage{graphicx}
\usepackage[usenames,dvipsnames]{color}
\usepackage{epstopdf}
\usepackage[all]{xy}
\usepackage[]{mcode}
\usepackage{supertabular}
\usepackage{subcaption}
\usepackage{array}

\providecommand{\U}[1]{\protect\rule{.1in}{.1in}}
\pdfoutput=1
\providecommand{\U}[1]{\protect\rule{.1in}{.1in}}

\newcommand{\qed}{\nobreak \ifvmode \relax \else
      \ifdim\lastskip<1.5em \hskip-\lastskip
      \hskip1.5em plus0em minus0.5em \fi \nobreak
      \vrule height0.75em width0.5em depth0.25em\fi}

\begin{document}

\title{Relay Assisted OFDM with Subcarrier Number Modulation in Multi-Hop Cooperative Networks}
\author{Shuping Dang, \textit{Member, IEEE}, Jiusi Zhou \textit{Student Member, IEEE}, Basem Shihada, \textit{Senior Member, IEEE}, and Mohamed-Slim Alouini, \textit{Fellow, IEEE}
  \thanks{S. Dang, J. Zhou, B. Shihada and M.-S. Alouini are with Computer, Electrical and Mathematical Science and Engineering Division, King Abdullah University of Science and Technology (KAUST), 
Thuwal 23955-6900, Kingdom of Saudi Arabia (e-mail: \{shuping.dang, jiusi.zhou, basem.shihada, slim.alouini\}@kaust.edu.sa).}}

\maketitle

\begin{abstract}
Orthogonal frequency-division multiplexing (OFDM) with subcarrier number modulation (OFDM-SNM) manifests its superior nature of high spectral efficiency (SE) and low complexity for signal estimation. To exploit the spatial gain of OFDM-SNM, we propose a relay assisted OFDM-SNM scheme for multi-hop cooperative systems in this letter. It is stipulated that a relay operated by decode-and-forward (DF) and half-duplex (HD) protocols exists in each hop. We analyze the outage performance of the relay assisted OFDM-SNM system. The average outage probability is approximated in closed form. Moreover, to reveal the diversity and coding gains of relay assisted OFDM-SNM, we explore the asymptotic outage performance at high signal-to-noise ratio (SNR) by power series expansion. To verify the improvement on outage performance and energy efficiency, we carry out the comparison among different multi-hops systems with traditional OFDM-SNM fixing a certain distance from source to destination. Simulation results corroborate the derived outage probabilities and provide insight into the proposed system in this letter.
\end{abstract}


\section{Introduction}
In recent years, spectrum scarcity has become one of the main barriers hindering the improvement of the quality of service (QoS) for wireless communications, which many researchers believe would soon lead to even a crisis on the corner \cite{6736752,dang2020should}. Index modulation (IM)  cooperating with multi-antenna systems and multi-carrier systems result in the well-known spatial modulation and orthogonal frequency-division multiplexing with IM (OFDM-IM), respectively. Both can effectively mitigate the spectrum scarcity problem \cite{8004416,8315127,8891788}. Spatial modulation requires multiple antennas and thereby multiple radio frequency (RF) chains, which will undoubtedly render an increased system complexity and physical size of communication device. Unfortunately, the requirements of high system complexity and large physical size of device might not be affordable by a number of new network paradigms where nodes are small, simple and power/complexity limited, e.g., wireless sensor networks (WSNs) \cite{7983347}. OFDM-IM suffers from a relatively high estimation complexity and difficulty of codebook design \cite{8353362,9003390}. In light of this challenge, OFDM with subcarrier number modulation (OFDM-SNM) was proposed, which is able to provide a higher spectral efficiency (SE) than OFDM-IM at a comparable level of error performance \cite{8362748,8703169,9042249}.

As a cognate alternative to OFDM-IM, OFDM-SNM employs the number of activated subcarriers to indicate the extra transmitted bit sequence instead of the indices of subcarriers. This results in a new \textit{number dimension} in addition to conventional signal amplitude and phase dimensions for signal transmission. On the other hand, unlike spatial modulation, OFDM-SNM suffers from the low energy efficiency and coverage without multiple antennas and accompanying beamforming functionality, which is also a major drawback with OFDM-IM. To mitigate the shortcomings in OFDM-IM, cooperative relaying is employed to collaborate with OFDM-IM, which yields the relay assisted OFDM-IM \cite{8075970,8358694,8476574,8730295}. Researchers have shown that by employing cooperative relaying, the performance of OFDM-IM can be improved by a considerable level, and the formed systems are easy to modify so as to involve various advanced techniques for transmission reinforcement, e.g., codebook optimization and relay selection \cite{8241721,8405601,8361430,8612925,8614439,8917612}.

Since cooperative relaying works very well with OFDM-IM, it naturally comes to the idea that we can also involve cooperative relaying to OFDM-SNM for performance enhancement. To concrete this idea into an actual form, we propose the relay assisted OFDM-SNM in this letter and analyze its outage performance. In particular, because the transmitted bit sequence by OFDM-SNM has a variable length, we approximate the average outage probability by averaging over all cases with different numbers of activated subcarriers. Furthermore, to reveal the diversity and coding gains of relay assisted OFDM-SNM, we utilize the power series expansion assuming that the ratio of transmit power to noise power becomes large and derive the asymptotic expression of average outage probability at high signal-to-noise ratio (SNR). All analytical results derived in this letter are substantiated by Monte Carlo simulations. We also discuss the impacts of a series of crucial parameters and settings on the outage performance by observing the numerical results from Monte Carlo simulations.

\section{System Model}\label{sm}

In this letter, a multi-hop relay assisted cooperative system is taken into consideration, where there is one source-destination pair connected by $L-1$ decode-and-forward (DF) relays in series. For simplicity, the direct transmission link and cross-hop transmission links between non-adjacent relays are omitted in this system. This is a common case for mobile cellular networks, e.g., WiMAX \cite{pejanovic2012ofdm}. To simplify the analysis and be representative, we focus on only a single group of $N$ subcarriers, which form the set $\mathcal{N}$. In general, through the inverse fast Fourier transform (IFFT), $N$ is normally a power of two in OFDM systems, though not necessarily. In OFDM-SNM, the purpose of employing these $N$ subcarriers is twofold. First, a subset $\mathcal{N}(k)$ of the $N$ subcarriers will be activated and the cardinality of the subset $\mathcal{N}(k)$ (i.e., the number of activated subcarriers) is denoted as $T(k)=\mathrm{Card}(\mathcal{N}(k))$, which is used for indicating the \textit{heading} bit stream $p_1$. $k$ is the index of the transmission pattern. The total number of transmission patterns in OFDM-SNM can be easily calculated by $\Xi=\sum_{\zeta=1}^{N}M^\zeta=\frac{M(M^N-1)}{M-1}$, where $M$ is the amplitude and phase modulation (APM) order. 

It is obvious that we have the relation between $p_1$ and $N$: $p_1=\lfloor\log_2(N)\rfloor$, where $\lfloor\cdot\rfloor$ is the floor function, which can be removed \textit{iff} $N$ is a power of two. Then, for each activated subcarrier, we can employ the $M$-ary phase-shift keying ($M$-PSK) to load APM constellation symbols on them, which are mapped from a $k$-dependent \textit{subsequent} bit stream with a variable length $p_2(k)=T(k)\log_2(M)$. Therefore, the entire variable-length bit stream for transmission is $p(k)=p_1+p_2(k)$, which is assumed to be equiprobable and uncorrelated in this letter. The average transmission rate in bit per channel use (bpcu) over all cases with different numbers of activated subcarriers can therefore be written as $\bar{p}=\underset{k}{\mathbb{E}}\left\lbrace p(k)\right\rbrace=\lfloor\log_2(N)\rfloor+\frac{N+1}{2}\log_2(M)$, where $\mathbb{E}\{\cdot\}$ denotes the expectation operation over the random variable enclosed.

To characterize a subcarrier activation pattern (SAP), we can involve the $k$-dependent activation state vector written as $\mathbf{S}(k)=[s(k,1),s(k,2),\dots,s(k,N)]^T\in\{0,1\}^{N\times 1}$, where $s(k,n)$ is a $k$-dependent binary indicator of the activation state of the $n$th subcarrier, which is completely determined by $p_1$. Then, with $\mathbf{S}(k)$ and $p_2(k)$, we can resort to IFFT by the standard OFDM scheme to yield an OFDM block for transmission purposes, which is given by $\mathbf{x}(k)=[x(k,1),x(k,2),\dots,x(k,N)]^T\in\mathbb{C}^{N\times 1}$, where $x(k,n)=\begin{cases}
\chi_n,~~~~\mathrm{if}~n\in\mathcal{N}(k)\\
0,~~~~~~\mathrm{otherwise}
\end{cases}$ and $\chi_n$ denotes the complex constellation symbol and for simplicity it is normalized by $\chi_n\chi_n^*=1$ without loss of generality.

\begin{figure}[!t]
\centering
\includegraphics[width=3.5in]{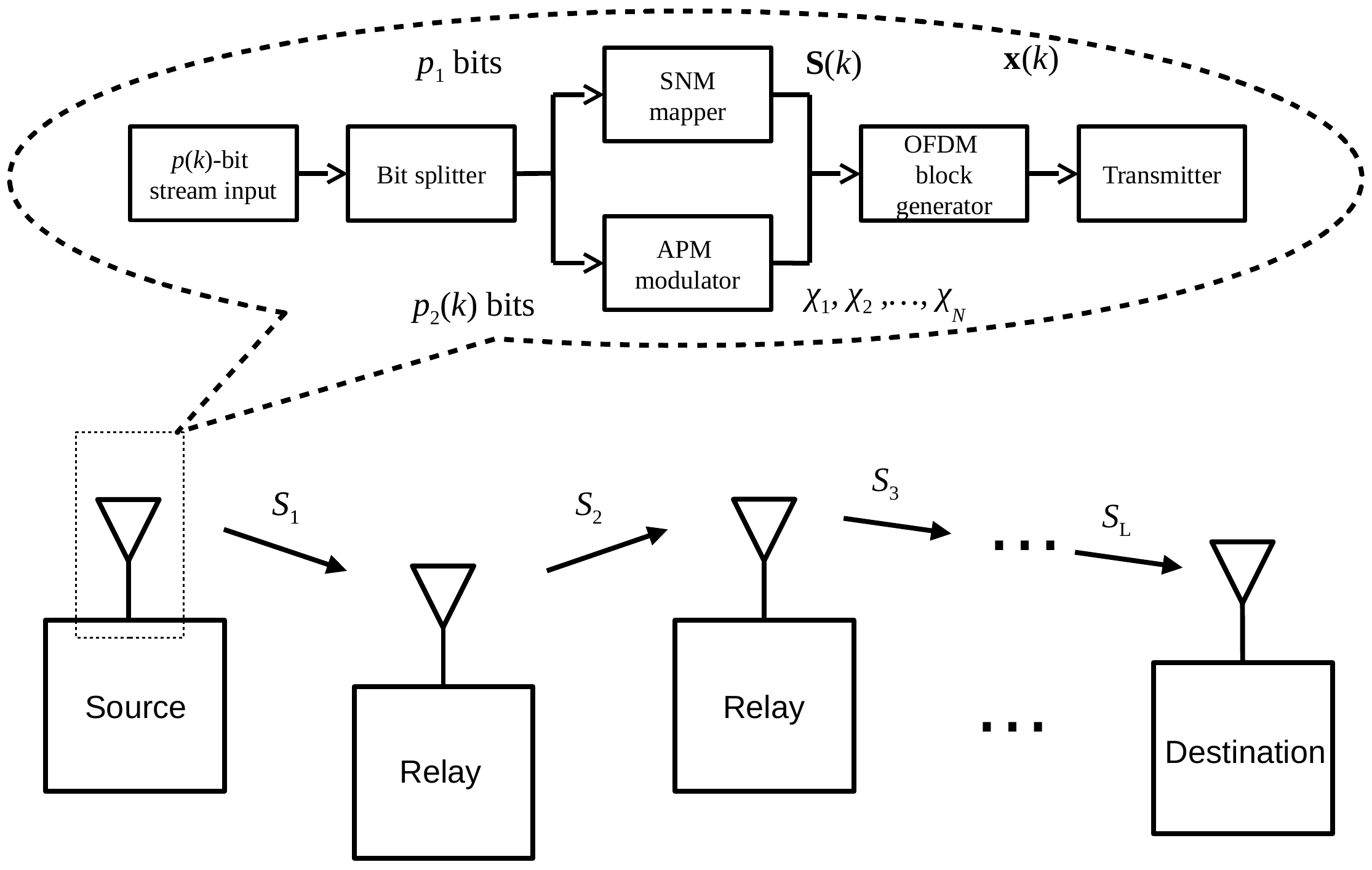}
\caption{Multi-hop relay assisted OFDM-SNM system and the transmitter design.}
\label{framework}
\end{figure}

Subsequently, the received OFDM block at the first DF relay is given by $\mathbf{y}_1(k)=\sqrt{\frac{P_t}{T(k)}}\mathbf{H}_1\mathbf{x}(k)+\mathbf{w}_1\in\mathbb{C}^{N\times 1}$, where $P_t$ is the aggregate transmit power at the source and relay nodes (assuming that they are homogeneous) and should be evenly allocated to $T(k)$ activated subcarriers; $\mathbf{H}_i=\mathrm{diag}\{h_i(1),h_i(2),\dots,h_i(N)\}\in\mathbb{C}^{N\times N}$ represents the channel state matrix for the $i$th hop, and $h_i(n)$ is the channel coefficient on the $n$th subcarrier, which is mutually independent $\forall~n\in\mathcal{N}$; $\mathbf{w}_i=[w_i(1),w_i(2),\dots,w_i(N)]^T\in\mathbb{C}^{N\times 1}$ represents the additive white Gaussian noise (AWGN) vector at the receiver (relay or destination) in the $i$th hop, and $w_i(n)\sim\mathcal{CN}(0,N_0)$ is the noise sample on the $n$th subcarrier and $N_0$ is the average noise power.

To benchmark the performance and provide the insight into relay assisted OFDM-SNM systems, the maximum-likelihood (ML) estimation method is utilized at all $L-1$ relays and the final destination to estimate the received OFDM block.  The ML estimation criterion for a received OFDM block at the receiver over the $i$th hop is given by $\hat{\mathbf{x}}_i(\hat{k}_i)=\underset{\dot{\mathbf{x}}(\dot{k})\in\mathcal{X}}{\arg\min}\begin{Vmatrix}{\mathbf{y}}_i({k}_i)-\sqrt{\frac{P_t}{T(\dot{k})}}\mathbf{H}\dot{\mathbf{x}}(\dot{k})\end{Vmatrix}_F$, where $\begin{Vmatrix}\cdot\end{Vmatrix}_F$ denotes the Frobenius norm of the enclosed matrix/vector; $\mathcal{X}$ is the set of all legitimate OFDM blocks in OFDM-SNM with the cardinality of $\mathrm{Card}(\mathcal{X})=\Xi$, which is the search space scale of OFDM block estimation and defines the estimation complexity. Therefore, for the receivers from the second to the last hops, the received OFDM block is given by $\mathbf{y}_i(\hat{k}_i)=\sqrt{\frac{P_t}{T(\hat{k}_i)}}\mathbf{H}_i\hat{\mathbf{x}}_i(\hat{k}_i)+\mathbf{w}_i\in\mathbb{C}^{N\times 1}$. We present the multi-hop relay assisted OFDM-SNM system and the transmitter design in Fig. \ref{framework} for clarity.

By normalizing the transmitted constellation symbol, we can derive the received SNR on the $n$th subcarrier at the receiver (relay or destination) in the $i$th hop to be\footnote{An exceptional notation is given for the first hop and thereby we denote $\hat{k}_0=k$ for simplicity.}
\begin{equation}\label{ueowjkawq2993}
\mathrm{SNR}_i(\hat{k}_{i-1},n)=\begin{cases}
\frac{P_tG_i(n)S_{i}^{-\alpha}}{T(\hat{k}_{i-1})N_0},~~~~n\in\mathcal{N}(\hat{k}_{i-1})\\
0,~~~~~~~~~~~~~~~~~\mathrm{otherwise}
\end{cases}
\end{equation}
where $S_i$ is the distance of the $i$th hop; $\alpha$ is the path loss exponent; $G_i(n)$ is the channel power gain pertaining to the $n$th subcarrier over the $i$th hop due to small-scale fading and is modeled as an identically distributed (i.i.d.) exponentially distributed random variable with unit mean. Therefore, the probability density function (pdf) and cumulative distribution function (cdf) are given by $f_i(g)=e^{-g}$ and $F_i(g)=1-e^{-g}$. We can easily have $\left|h_i\left(n\right)\right|^2 = G_i\left(n\right)S_i^{-\alpha}$.

\section{Outage Performance Analysis}\label{opa}
To simplify the performance analysis, we simply assume that as long as the subcarrier-wise received SNR at the receiver is larger than a predetermined outage threshold $\xi$, the constellation symbol conveyed on that subcarrier can be successfully decoded and re-transmitted, which refers to the DF relaying protocol. Meanwhile, we assume a half-duplex transmission mechanism is adopted at all DF relays, which indicates that an end-to-end transmission from source to destination through $L-1$ DF relays in series requires $L$ orthogonal time slots. With both assumptions, we can decompose the outage performance analysis in terms of subcarrier and hop and therefore simplify the derivation of the end-to-end outage probability.
\subsection{Definition of End-to-End Performance Metric}\label{daop}
To evaluate the system reliability in a comprehensive manner, we first formulate the subcarrier-wise per-hop conditional outage probability on SAP $\hat{k}_{i-1}$ for the $n$th subcarrier in the $i$th hop, which is the probability of the random event that $\mathrm{SNR}(\hat{k}_{i-1},n)<\xi$, $\forall~n\in\mathcal{N}(\hat{k}_{i-1})$. Mathematically, the subcarrier-wise per-hop conditional outage probability on $\hat{k}_{i-1}$ is given by
\begin{equation}\label{tengs98721}
\Phi_i(\hat{k}_{i-1},n)=\mathbb{P}\left\lbrace \mathrm{SNR}_i(\hat{k}_{i-1},n)<\xi\right\rbrace,
\end{equation}
where $\mathbb{P}\left\lbrace\cdot\right\rbrace$ represents the probability of the random event included.

In general, the information carried on activated subcarriers has specific relations for error detection, error correction, and synchronization. They pose a stringent demand on multi-carrier signal estimation. It stipulates that all correctly activated subcarriers have to be detected with SNRs exceeding the predetermined outage threshold $\xi$ \cite{6157252}. Meanwhile, because of the bottleneck effect and error propagation for DF cooperative relaying, we formulate the end-to-end conditional outage probability over $L$ hops in (\ref{changchangdeoutageprobdingy}) at the top of the next page. Furthermore, to consider all cases with different SAPs, we give an explicit definition of the average outage probability by averaging $\Phi(k)$ over SAP $k$ and have $\overline{\Phi}=\underset{k}{\mathbb{E}}\left\lbrace\Phi(k)\right\rbrace$. This metric is employed in this letter to characterize the system reliability of the proposed relay assisted multi-hop OFDM-SNM system.

\begin{figure*}[!t]
\begin{equation}\label{changchangdeoutageprobdingy}
\begin{split}
\Phi(k)=\mathbb{P}\left\lbrace\left\lbrace\underset{i\in\{1,2,\dots,L\}}{\bigcup}\left\lbrace\underset{n\in\mathcal{N}(\hat{k}_{i-1})}{\bigcup}\left\lbrace\mathrm{SNR}_i(\hat{k}_{i-1},n)<\xi\right\rbrace\right\rbrace\right\rbrace\bigcup\left\lbrace \underset{i\in\{1,2,\dots,L-1\}}{\bigcup}\left\lbrace\hat{k}_i\neq k\right\rbrace\right\rbrace\right\rbrace
\end{split}
\end{equation}
\hrule
\end{figure*}

\subsection{Approximation}
To determine the average outage probability formulated in the last subsection, we can start from the very beginning by deriving the subcarrier-wise per-hop conditional outage probability on SAP $k$ according to (\ref{ueowjkawq2993}) and (\ref{tengs98721}) to be
\begin{equation}\label{dasj22uiks222}
 \begin{split}
 \Phi_i(\hat{k}_{i-1},n)&=\mathbb{P}\left\lbrace G_i(n)<\frac{T(\hat{k}_{i-1})N_0\xi}{P_tS_i^{-\alpha}}\right\rbrace\\
&=F_i\left(\frac{T(\hat{k}_{i-1})N_0\xi}{P_tS_i^{-\alpha}}\right).
\end{split}
\end{equation}
In order to facilitate the following analysis and obtain an insightful expression, we make an assumption that as long as all SNRs regarding the subset of activated subcarriers in the $i$th hop are larger than the predetermined threshold $\xi$, the activation pattern index $\hat{k}_i$ can be correctly detected. This assumption reduces (\ref{changchangdeoutageprobdingy}) to be
\begin{equation}
\begin{split}
\Phi(k)\approx\mathbb{P}\left\lbrace\underset{i\in\{1,2,\dots,L\}}{\bigcup}\left\lbrace\underset{n\in\mathcal{N}(k)}{\bigcup}\left\lbrace\mathrm{SNR}_i(k,n)<\xi\right\rbrace\right\rbrace\right\rbrace.
\end{split}
\end{equation}
Subsequently, we can approximate $\Phi(k)$ in terms of $\Phi_i(k,n)$ given $\hat{k}_1=\hat{k}_2=\dots=\hat{k}_{L-1}=k$ by fundamental relations in probability theory as
\begin{equation}\label{closj22222daska}
\begin{split}
\Phi(k)&\approx 1-\prod_{i\in\{1,2,\dots,L\}}\left[\prod_{n\in\mathcal{N}(k)}\left(1-\Phi_i(k,n)\right)\right]\\
&\overset{(\mathrm{a})}{=}1-\prod_{i\in\{1,2,\dots,L\}}\left(1-\Phi_i(k,n)\right)^{T(k)},
\end{split}
\end{equation}
where $(\mathrm{a})$ is justified by the homogeneity of subcarrier assumed in this letter.

Finally, according to the mapping relation between incoming bit stream and SAP stipulated in Section \ref{sm}, we can easily derive the average outage probability by $\overline{\Phi}=\underset{k}{\mathbb{E}}\left\lbrace\Phi(k)\right\rbrace$ to be
\begin{equation}\label{tajssss21931023}
\overline{\Phi}=\sum_{\zeta=1}^{N}\Upsilon(\zeta)\Phi(k)\vert_{T(k)=\zeta}=\frac{1}{N}\sum_{\zeta=1}^{N}\Phi(k)\vert_{T(k)=\zeta},
\end{equation}
where $\Upsilon(\zeta)=\mathbb{P}\left\lbrace T(k)=\zeta\right\rbrace=1/N$ represents the probability that $\zeta$ subcarriers are activated, which is resulted by the equiprobable and uncorrelated $p(k)$-bit stream.

\subsection{Asymptotic Analysis}
To reveal the diversity and coding gains of relay assisted OFDM-SNM and provide insightful information, the asymptotic expression for average outage probability in the high SNR region ($P_t/N_0\rightarrow\infty$) is derived as follows by wielding power series expansion in this subsection. According to (\ref{dasj22uiks222}), it is evident that when $P_t/N_0\rightarrow\infty$, $\Phi_i(k,n)\rightarrow 0$. By this property, we can first reduce the approximate end-to-end conditional outage probability $\Phi(k)$ given in (\ref{closj22222daska}) at high SNR to be
\begin{equation}\label{phisaskda2}
\Phi(k)\approx T(k)\sum_{i\in\{1,2,\dots,L\}} \Phi_i(k,n),
\end{equation}
and the detailed proof is shown in Appendix \ref{appendix:A}. By power series expansion, the subcarrier-wise per-hop conditional $\Phi_i(k,n)$ can be expanded at high SNR as $\Phi_i(k,n)\sim \frac{T(k)N_0\xi}{P_t S_i^{-\alpha}}$, which can be substituted into (\ref{phisaskda2}) to yield $\Phi(k)\sim \frac{T(k)^2N_0\xi}{P_t S_{\Sigma}}$, where $S_{\Sigma}=\frac{1}{\sum_{i=1}^{L}{S_i^{\alpha}}}$ is the average end-to-end channel power gain. Finally, substituting the asymptotic expression of $\Phi(k)$ into (\ref{tajssss21931023}) produces a neat asymptotic expression for the average outage probability in the high SNR region:
\begin{equation}\label{longlong}
\begin{split}
\overline{\Phi}&\sim\frac{N_0\xi(N+1)(2N+1)}{6P_tS_{\Sigma}}.
\end{split}
\end{equation}

\section{Numerical Results and Discussions}\label{nrd}
To corroborate the analysis and discover the proposed relay assisted OFDM-SNM system, we present and discuss the simulation results generated by Monte Carlo methods in this section. In the first place, we initialize the simulation platform by setting up system parameters. Here, for the purposes of verifying the improvement of relay assisted multi-hop OFDM-SNM, we specify two application scenarios to study the outage performance and energy efficiency respectively. We fix the total transmission distance between source node and destination via different numbers of relay nodes by $\Sigma_{i}S_i=5$ and set $\alpha = 2$ and $\xi=1$. In this first case, we consider the same ratio of transmit power to noise power at all transmitters of the source and relays to investigate outage performance. In the second case, we uniformly allocate the ratio of transmit power to noise power to all transmitters to facilitate the analysis of energy efficiency.

\begin{figure}[!t]
\centering
\includegraphics[width=3.2in]{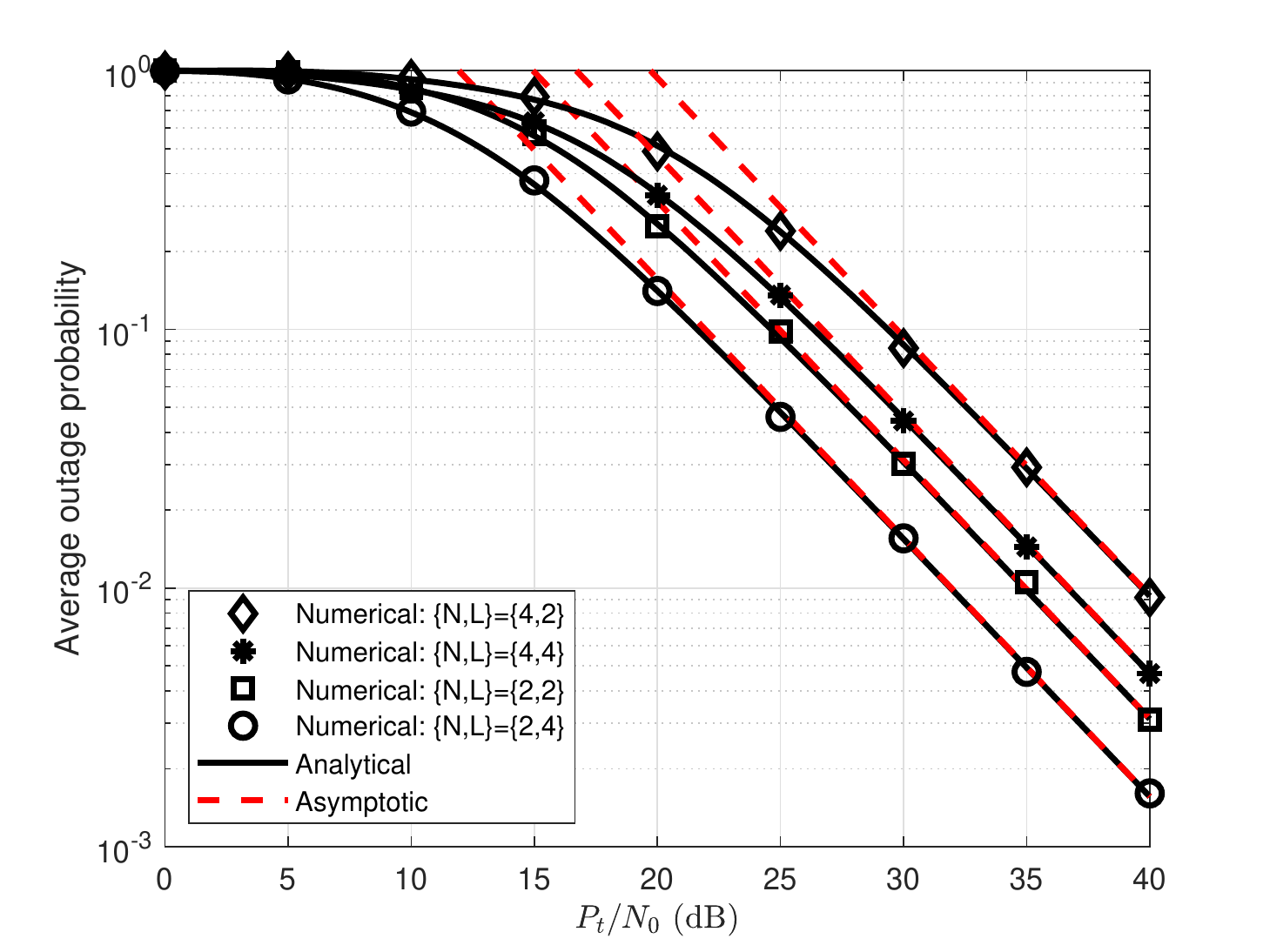}
\caption{Relation between the average outage probability and $P_t/N_0$, given the same transmit power at all  nodes.}
\label{outage_Pt}
\end{figure}

We first vary the target parameter $P_t/N_0$ with various numbers of subcarriers and hops to substantiate the analytical results regarding outage performance derived in (\ref{tajssss21931023}) and (\ref{longlong}) when binary PSK is in use. The simulation results are illustrated in Fig. \ref{outage_Pt}. By observing Fig. \ref{outage_Pt}, most importantly, we can corroborate the analytical results pertaining to outage performance derived in (\ref{tajssss21931023}) by the perfectly matched numerical and analytical curves. Also, the asymptotic curves can also well match the numerical curves at a moderate range of SNR, which testifies the feasibility of (\ref{longlong}). Meanwhile, all cases have the same diversity order due to the one-diversity system model stipulated in this letter, but different numbers of subcarriers and hops $N$ and $L$ will lead to a considerable coding shift. 

\begin{figure}[!t]
    \centering
    \begin{subfigure}[t]{0.5\textwidth}
        \centering
        \includegraphics[width=3.2in]{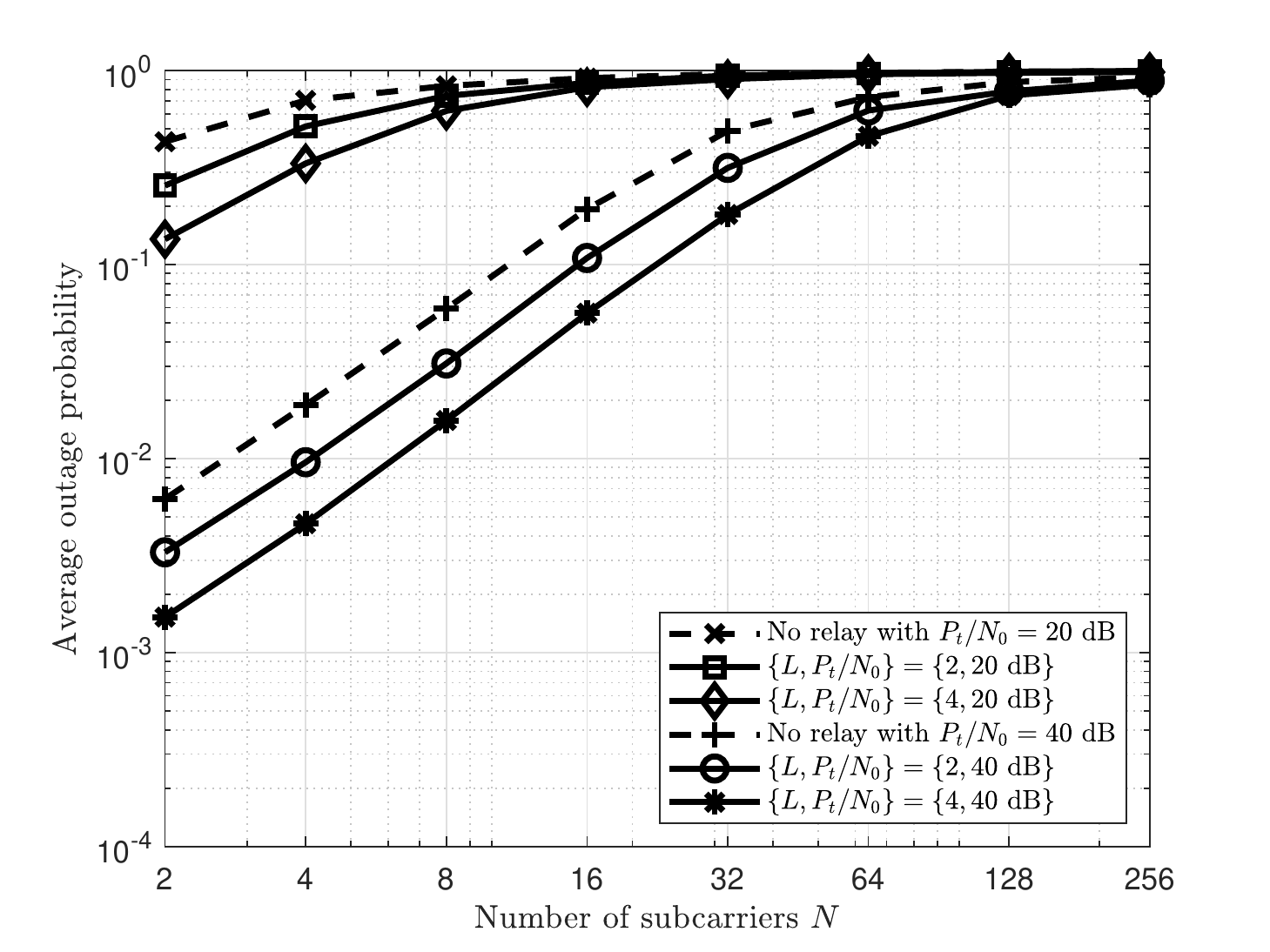}
        \caption{Number of subcarriers $N$}
    \end{subfigure}%

    \begin{subfigure}[t]{0.5\textwidth}
        \centering
        \includegraphics[width=3.2in]{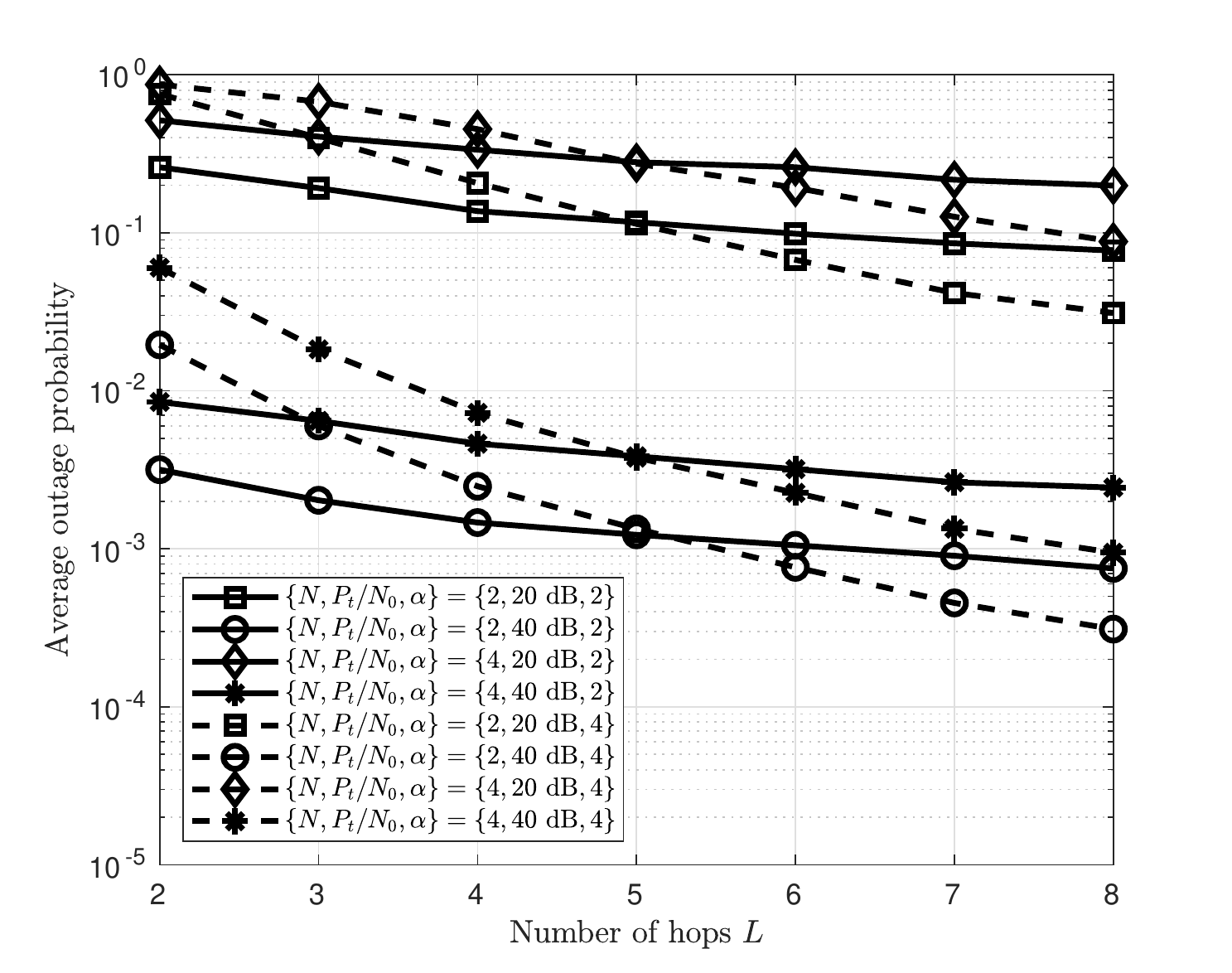}
        \caption{Number of hops $L$}
    \end{subfigure}
    \caption{Relation between the average outage probability and crucial system parameters, given the same transmit power at all nodes.}
    \label{outage_NL}
\end{figure}

Concerning the effects of $N$ and $L$, further investigations are worthwhile. The relation between average outage probability and $N$ as well as $L$ is simulated under the same configurations as for Fig. \ref{outage_Pt} and shown in Fig. \ref{outage_NL}. From this figure, it is evident that increasing $N$ or decreasing $L$ will lead to a higher average outage probability. Meanwhile, Fig. \ref{outage_NL} illustrates the outage performance comparison with equal transmit power over all nodes, which verifies the improvement on outage performance by introducing multi-hop relaying to OFDM-SNM. On the other hand, with an increasing number of relays, the brought improvement on outage performance gradually decreases. This forms a trade-off between the improvement on outage performance and relay implementation cost. This trade-off is of particular importance for practical relay assisted OFDM-SNM systems and worth investigating as future work.

\begin{figure}[!t]
\centering
\includegraphics[width=3.0in]{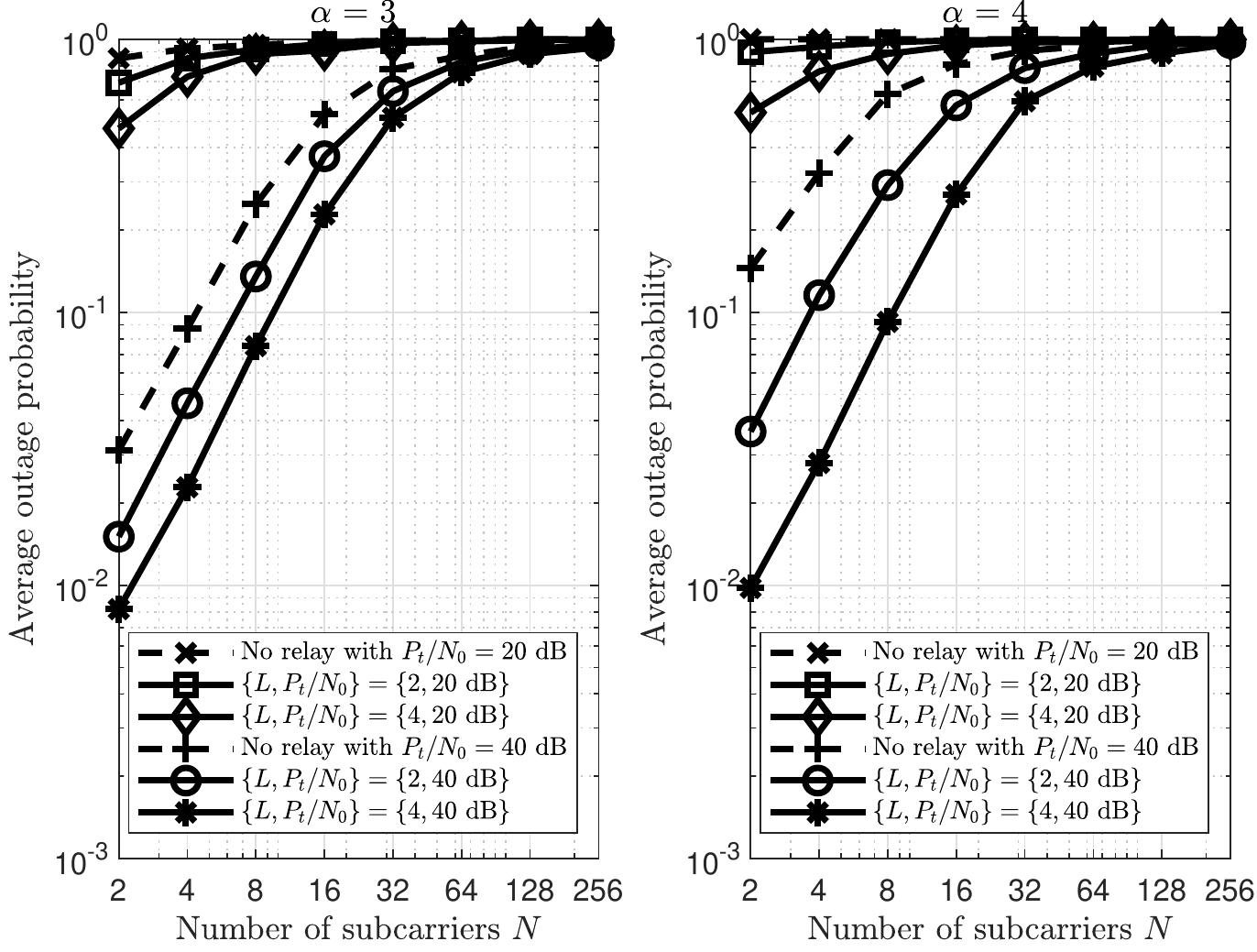}
\caption{Relation between the average outage probability and numbers of subcarriers $N$ for different path loss exponents, given the total transmit power uniformly allocated to all transmitting nodes.}
\label{outage_alpha}
\end{figure}

Then, with different $\alpha$, we simulate the proposed system by fixing the total ratio of transmit power to noise power and allocating it uniformly to all transmitters to explore the energy efficiency in Fig. \ref{outage_alpha}. It is clear that the energy efficiency is also enhanced by introducing multi-hop relaying, as increasing the number of hops results in a lower average outage probability.

\section{Conclusion}\label{c}
In this letter, we proposed a relay assisted multi-hop OFDM-SNM scheme and analyzed its outage performance. Specifically, we approximated the average outage probability of the proposed system in closed form, which is defined in a comprehensive manner capturing the unique features of multi-carrier and multi-hop communications. In addition, we performed power series expansion on the approximate expression and obtained the asymptotic expression in at high SNR. All analyses have been validated by the simulation results. Based on simulation results, we also discussed the effects of the numbers of subcarriers and hops on the outage performance. Based on the comparison with traditional OFDM-SNM without the support of relay, we verified the outage performance improvement brought by the multi-hop architecture. Besides, by fixing the total transmit power used throughout the entire network, we also verified the improvement on energy efficiency by introducing multi-hop relaying to OFDM-SNM. On the other hand, it is admitted that introducing such a multi-hop relaying architecture inevitably raises the system complexity. In order to measure the incurred complexity and study the performance-complexity trade-off, a cross-layer system model of multi-hop relay assisted OFDM-SNM should be constructed, which is worth investigating as future work.

\appendices

\section{} \label{appendix:A}

By the approximation given in (\ref{closj22222daska}), we can easily derive the asymptotic expression of $\Phi(k)$ when $\Phi_i(k,n)\rightarrow 0$ to be

\begin{equation}\label{appendixA} \small
\begin{split} 
\Phi(k)&= 
{1-\prod_{i\in\{1,2,\dots,L\}}\left(1-\Phi_i(k,n)\right)^{T(k)}}\\
&{=} 1-\prod_{i\in\{1,2,\dots,L\}}\left[\sum_{j=0}^{T(k)}\binom{T(k)}{j}\left(-\Phi_i(k,n)\right)^{j} 1^{T-j}\right]\\
&{\overset{(\mathrm{b})}\approx} 1-\prod_{i\in\{1,2,\dots,L\}}\left[\sum_{j=0}^{1}\binom{T(k)}{j}\left(-\Phi_i(k,n)\right)^{j} 1^{T-j}\right]\\
&{=}1-\prod_{i\in\{1,2,\dots,L\}}\left(1-T(k)\Phi_i(k,n)\right)\\
&{\overset{(\mathrm{c})}\approx}1-\left[1-T(k)\Phi_1(k,n)-\cdots-T(k)\Phi_L(k,n)\right]\\
&{=}T(k)\sum_{i\in\{1,2,\dots,L\}} \Phi_i(k,n),
\end{split}
\end{equation}
where (b) is approximated by keeping the two dominant terms of the binomial expansion since $\Phi_i(k,n)\rightarrow 0$; (c) is derived by keeping the  first order terms of $\Phi_i(k,n)$.

%\section*{Acknowledgment}
%The authors thank the editor and the anonymous reviewers for their constructive comments which help us improve the quality of this paper.

\bibliographystyle{IEEEtran}
\bibliography{bib}

\end{document}